\documentclass[aps,prd,superscriptaddress,showpacs,nofootinbib,10pt,notitlepage,colorlinks=true,allcolors=BlueViolet,hyperfootnotes=false,preprintnumbers]{revtex4-2}

\usepackage{graphicx}
\usepackage{amsmath,amssymb}
\usepackage{bbm}
\usepackage[dvipsnames]{xcolor}
\usepackage[utf8]{inputenc}
\usepackage{lmodern}
\usepackage[T1]{fontenc}
\usepackage[inline]{enumitem}
\usepackage{accents}
\usepackage{slashed}
\usepackage{orcidlink}
\usepackage{natbib}
\usepackage{stackengine}
\usepackage{soul}
\usepackage{braket}

\usepackage[normalem]{ulem}

\usepackage{tikz}
\usetikzlibrary{shapes.geometric}

\definecolor{ao(english)}{rgb}{0.0, 0.5, 0.0}

\newcommand{\olsi}[1]{\,\overline{\!{#1}}} 

\newcommand{\be}{\begin{equation}} \newcommand{\ee}{\end{equation}}
\newcommand{\ba}{\begin{array}{c}} \newcommand{\ea}{\end{array}}
\newcommand{\bea}{\begin{eqnarray}} \newcommand{\eea}{\end{eqnarray}}

\graphicspath{{Figures/}}

\begin{document}


\title{\boldmath Melting down a tetraquark: \texorpdfstring{$D^{\ast}D^{(\ast)}$}{D*D(*)} interactions and \texorpdfstring{$T_{cc}(3875)^+$}{Tcc(3875)+} in a hot environment}
\preprint{JLAB-THY-25-4398}

\newcommand{\ific}{Instituto de F\'{\i}sica Corpuscular (centro mixto CSIC-UV),
Institutos de Investigaci\'on de Paterna,
C/Catedr\'atico Jos\'e Beltr\'an 2, E-46980 Paterna, Valencia, Spain}
\newcommand{\ice}{Institute of Space Sciences (ICE, CSIC), Campus UAB,  Carrer de Can Magrans, 08193 Barcelona, Spain}
\newcommand{\ieec}{Institut d'Estudis Espacials de Catalunya (IEEC), 08860 Castelldefels (Barcelona), Spain}
\newcommand{\jlab}{Theory Center, Thomas Jefferson National Accelerator Facility, Newport News, VA 23606, USA}

\author{V.~Montesinos\orcidlink{0000-0002-6186-2777}}
\email{Victor.Montesinos@ific.uv.es}
\affiliation{\ific}
\author{G. Montana\orcidlink{0000-0001-8093-6682}}
\email{gmontana@jlab.org}
\affiliation{\jlab}
\author{M.~Albaladejo\orcidlink{0000-0001-7340-9235}}
\email{Miguel.Albaladejo@ific.uv.es}
\author{J.~Nieves\orcidlink{0000-0002-2518-4606}}
\email{jmnieves@ific.uv.es}
\affiliation{\ific}
\author{L.~Tolos\orcidlink{0000-0003-2304-7496}}
\email{tolos@ice.csic.es}
\affiliation{\ice}
\affiliation{\ieec}

\begin{abstract} We discuss the modification of the properties of the tetraquark-like $T_{cc}(3875)^+$ and its heavy quark spin partner, $T_{cc}(4016)^{*+}$ immersed in a hot bath of pions. We consider these exotic states as purely isoscalar $DD^\ast$ and $D^\ast D^\ast$ $S$-wave bound states, respectively. Finite temperature effects are incorporated through the $D$ and $D^\ast$ state-of-the-art thermal spectral functions calculated in [G. Montana \textit{et al.}, \textit{Phys. Rev. D}, \textbf{102} (2020) 096020], using the imaginary-time formalism. We find important modifications of the $DD^\ast$ and $D^\ast D^\ast$ scattering amplitudes already for $T=80$ MeV, and show that the hot-bath lineshapes of these tetraquark-like states strongly depend on their Weinberg molecular content. We find that the thermal $T_{cc}(3875)^+$ and $T_{cc}(4016)^{*+}$ spectral functions change more rapidly with temperature for high molecular probabilities $P_0$. For large values of $P_0$, the widths significantly increase with temperature, leading to the melting of these exotic states for temperatures larger than 80 MeV. For small molecular components, the changes in the spectral functions of these states due to temperature become significantly less important. All these results show that any future experimental determination of the $D^{(\ast)}D^*$ scattering amplitudes at finite temperature will provide valuable insights into the molecular content of the $T_{cc}(3875)^+$ and $T_{cc}(4016)^{*+}$ exotics.
\end{abstract}

\maketitle

\section{Introduction}
Since the beginning of the century, several new hadronic states have been experimentally determined. Starting from the pioneer discovery of the $X(3872)$ in 2003 by the Belle Collaboration \cite{Belle:2003nnu}, a plethora of charmonium-like states have been reported, triggering the interest of the hadronic community (see some recent reviews in \cite{Guo:2017jvc,Liu:2019zoy,Brambilla:2019esw,Dong:2021bvy,Chen:2022asf}). Among them, the recently discovered $T_{cc}(3875)^+$  \cite{LHCb:2021vvq,LHCb:2021auc} is taking a prominent role. Observed in the $D^0D^0 \pi^+$ mass distribution, this state has a mass of $m_{\rm{thr}} + \delta m_{\rm{exp}}$, with $m_{\text{thr}} = 3875.09\,\text{MeV}$ being the $D^{*+} D^0$ mass threshold and $\delta m_{\rm{exp}} = -360 \pm 40^{+4}_{-0}\,\text{keV}$, whereas its width is $\Gamma = 48 \pm 2^{+0}_{-14}\,\text{keV}$ \cite{LHCb:2021auc}.

From the theoretical perspective, there has been an extensive discussion on the nature of the $T_{cc}(3875)^+$ state. On the one hand, several works advocate for the molecular interpretation of this state \cite{Janc:2004qn,Carames:2011zz,Ohkoda:2012hv,Li:2012ss,Liu:2019stu,Dong:2021bvy,Feijoo:2021ppq,Ling:2021bir,Fleming:2021wmk,Ren:2021dsi,Chen:2021cfl,Albaladejo:2021vln,Du:2021zzh,Baru:2021ldu,Santowsky:2021bhy,Deng:2021gnb,Ke:2021rxd,Agaev:2022ast,Meng:2022ozq,Abreu:2022sra,Chen:2022vpo,Albaladejo:2022sux,Dai:2023cyo, Wang:2023ovj} due to its proximity to the $D^0D^{*+}$ and $D^+D^{*0}$ thresholds. On the other hand, the tetraquark description has been assumed, even before its discovery \cite{Ballot:1983iv,Zouzou:1986qh,Navarra:2007yw,Ebert:2007rn,Karliner:2017qjm,Yang:2019itm}. However, the proximity to the $D^0D^{*+}$ and $D^+D^{*0}$ thresholds leads to considering hadronic degrees of freedom for the correct analysis of the experimental data \cite{Dong:2020hxe,Dong:2021juy,Dai:2023kwv}. 

Given the interest in the $T_{cc}(3875)^+$ state, ongoing work aims to identify scenarios in which its nature and properties can manifest more clearly. Recent efforts have been made to generate and understand the nature of the $T_{cc}(3875)^+$ state using lattice QCD results \cite{Padmanath:2022cvl,Chen:2022vpo,Lyu:2023xro,Collins:2024sfi,Whyte:2024ihh, Du:2023hlu, Gil-Dominguez:2024zmr}. Also, work on femtoscopic correlation functions has been recently performed for the $T_{cc}(3875)^+$ state in Refs.~\cite{Kamiya:2022thy,Vidana:2023olz,Albaladejo:2023wmv}.  

Another possible way to learn about the nature of the $T_{cc}(3875)^+$ is to analyze its behavior under the extreme conditions of density and/or temperature present at Relativistic Heavy Ion Collider (RHIC), Large Hadron Collider (LHC) or the future Facility for Antiproton and Ion Research (FAIR) energies. In Ref.~\cite{Fontoura:2019opw}, the production of exotic tetraquarks, such as $T_{QQ}$ with $Q=c,b$, was investigated using the quark coalescence model for Pb$+$Pb collisions at the LHC, showing production yields one order of magnitude smaller than previous estimates. In Ref.~\cite{Hu:2021gdg}, the centrality dependence, rapidity, transverse momentum and elliptic flow in Pb$+$Pb for LHC energies for $T_{cc}(3875)^+$ (as well as its potential isospin partners) were analyzed within the molecular picture using the AMPT transport model including coalescence. The study found a strong enhancement of the $T_{cc}$ yield in Pb$+$Pb collisions relative to $pp$, comparable to the $X(3872)$ one in central collisions, while showing a considerably stronger decrease toward peripheral events. Also, in Ref.~\cite{Yun:2022evm} the $T_{cc}$ was investigated within the coalescence model in Pb$+$Pb at 5.02 TeV, concluding that it could either be a compact multi-quark configuration or a loosely bound molecular state composed of charmed mesons. 

More recently, other production mechanisms have been proposed to understand the nature of this exotic state, such as photoproduction reactions involving one or two photons processes in ultra-peripheral collisions \cite{Wang:2023zay}, prompt production of the $T_{cc}^+$ (and its antimatter counterpart $T_{\bar c \bar c}^-$) in $pp$ collisions at a center-of-mass energy of 14 TeV \cite{Hua:2023zpa}, photoproduction off nuclei at near-threshold photon beam energies of 30–38 GeV that could be accessible in the proposed high-luminosity Electron-Ion Collider (EiC) and Electron-Ion Collider in China (EicC) \cite{Paryev:2024ors}, indirect production mechanisms through high-energy decay processes involving Higgs, $Z_{0}$ and $W^+$ at LHC or the Circular Electron-Positron Collider (CEPC) \cite{Niu:2024ghc}, and  photoproduction via photon-gluon fusion at  the International Linear Collider (ILC) and the Compact Linear Collider (CLIC) \cite{Niu:2025gcj}.

With the objective of analyzing the finite-density regime generated at the CBM experiment at FAIR, in Ref.~\cite{Montesinos:2023qbx} we have addressed the properties of $T_{cc}(3875)^+$ and  $T_{ \bar c \bar c}(3875)^-$ as well as their heavy-quark spin partners in nuclear matter, highlighting the distinctive density pattern of this particle-antiparticle pair if a small or a large molecular component in these tetraquark-like states was assumed.\footnote{A dense nuclear medium induces also sizable charge-conjugation asymmetries in the $D_{s0}^\ast(2317)^\pm$ and $D_{s1}(2460)^\pm$ resonances, seen as isoscalar $D^{(*)}K$ and $\olsi{D}{}^{(*)}\olsi{K}$ $S$-wave bound states, mainly due to the very different kaon and antikaon interactions with nuclear matter. In Ref.\,\cite{Montesinos:2024uhq}, we discussed in detail how this new feature can be used to better determine/limit the internal structure of these exotic states, similar to what was done in Ref.~\cite{Montesinos:2023qbx} for the $T_{cc}(3875)^+$ and $T_{\bar c \bar c}(3875)^-$.} In the present work, we aim to study the behavior of the $T_{cc}(3875)^+$ at high temperatures, such as those produced at RHIC or LHC. Owing to heavy-quark spin symmetry (HQSS), the $T_{cc}(3875)^+$ has a plausible symmetric partner in the $I(J^P) = 0(1^+)$ $D^\ast D^\ast$ channel, the $T_{cc}(4016)^{*+}$. This state has been studied in several theoretical works as dynamically generated~\cite{Albaladejo:2021vln,Du:2021zzh,Dai:2021vgf,Montesinos:2023qbx}, but has not yet been experimentally confirmed. Our approach allows us to investigate this HQSS partner at finite temperature as well. 

The present work builds upon our earlier finite-temperature studies of open-charm mesons~\cite{Montana:2020lfi,Montana:2020vjg,MontanaFaiget:2022cog} and their extension to the $X(3872)$~\cite{Montana:2022inz}. Applications of the results include the calculation of transport coefficients for heavy mesons in the hadronic phase~\cite{Torres-Rincon:2021yga}, and to the study of open-charm meson Euclidean correlators for the direct comparison with lattice QCD data~\cite{Montana:2020var}. While the main objective of this letter is to provide a qualitative description of the in-medium modifications and melting of the $T_{cc}$ and $T_{cc}^\ast$ states, the results presented here set the foundations for future finite-temperature analyses, including potential comparisons with lattice QCD via Euclidean correlators. In addition, the in-medium scattering amplitudes and spectral functions obtained with our framework may serve as microscopic input for future heavy-ion collision simulations, enabling connections to experimental observables.

In our approach, the $T_{cc}(3875)^+$ and $T_{cc}(4016)^{*+}$ are generated as bound states from the interactions of the $DD^*$ and $D^*D^*$ mesons, respectively, assuming either a large or a small molecular component. The changes in the $D$ and $D^*$ propagators at finite temperature are implemented via the Imaginary-Time Formalism (ITF) so as to finally obtain the  $T_{cc}(3875)^+$ and $T_{cc}(4016)^{*+}$ thermal scattering amplitudes and spectral functions. The $T_{ \bar c \bar c}(3875)^-$ and $T_{ \bar c \bar c}(4016)^{*-}$ at finite temperature behave similarly as $T_{cc}(3875)^+$ and $T_{cc}(4016)^{*}$, respectively, given that the constituents $\olsi{D}$ and $\olsi{D}{}^{\ast}$ interact equally with a hot thermal bath of light mesons as the $D$ and $D^*$.

The paper is organized as follows. In Sec.~\ref{sec:formalism} we present the $D^{(\ast)}D^*$ scattering amplitudes and the dynamical generation of the $T_{cc}(3875)^+$ in free space and at finite temperature.  We employ the ITF to obtain the two open-charm-meson thermal propagator, which determines the scattering amplitudes when the $D$ and $D^*$ mesons are immersed in the hot pion bath. In Sec.\,\ref{sec:results} we present our results  for the $T_{cc}(3875)^+$ (Subsec.\,\ref{ss:resultsTcc}), as well as  for the $T_{c c}^*(4016)^+$  (Subsec.\,\ref{ss:resultsheavy-quark}). The main conclusions of our work are given in Sec.\,\ref{sec:conclusions}.

\section{Formalism}
\label{sec:formalism}

We start by considering the $T_{cc}(3875)^+$ as an $S$-wave $DD^*$ state with isospin and spin-parity quantum numbers $I(J^P) = 0(1^+)$. Following our previous works~\cite{Albaladejo:2021cxj,Montesinos:2023qbx,Montesinos:2024uhq}, we take into account two families of energy-dependent contact potentials, which are expanded around threshold as:
\begin{subequations}
\label{e:Potential}
\begin{align} 
   V_A(s) &= C_1 + C_2\, [s-(m_D+m_{D^*})^2]\,, \label{eq:VA}\\
    V_B(s) &= \left(C_1^\prime + C_2^\prime\,  [s-(m_D+m_{D^*})^2]\right)^{-1}\,, \label{eq:VB}
\end{align}  
\end{subequations}
where $s=P^2$, with $P^\mu$ the total four-momentum of the $DD^*$ pair. The quantities $C_1^{(\prime)}$ and $C_2^{(\prime)}$ are low-energy constants (LECs) that have to be adjusted, as we show in the following. 

These interactions result from retaining the first two orders of the Taylor expansion around $s=m_D+m_{D^*}$, either of the potential $V(s)$ (type $A$) or of the inverse of the potential $V^{-1}(s)$ (type $B$). Note that the $V_A(s)$ potential contains constant terms that give rise to purely molecular states as well as some contributions related to the exchange of genuine compact quark-model structures, whereas the $V_B(s)$ is generated by the exchange of the bare quark-model $T_{cc}(3875)^+$ state (see Ref.~\cite{Montesinos:2023qbx} for the discussion).   
With these potentials, we solve the Bethe-Salpeter Equation (BSE) for the $T$-matrix within the on-shell approximation~\cite{Nieves:1999bx},
\be
{\cal T}^{-1}(s)= V^{-1}(s)-\Sigma_0(s), 
\ee
where $\Sigma_0(s)$ is the $D D^*$ loop function in the vacuum given by
\begin{align}
    \Sigma_0(s) = i \int \frac{d^4q}{(2\pi)^4} \Delta_D(P-q)  \Delta_{D^*}(q), \\
    \Delta_{D,D^*}(q) = \frac{1}{(q^0)^2-\vec{\,q}^{\,2} - m_{D,D^*}^2 + i\varepsilon},
\end{align}
which requires introducing an ultraviolet (UV) regulator in the $d^3q$ integration to make
the two-point function $\Sigma_0(s)$ finite. In this work, we will use a sharp momentum cutoff, $\Lambda=0.7$ GeV, as previously done in Refs.~\cite{Montesinos:2023qbx,Montesinos:2024uhq}.

To determine the LECs of the $DD^*$ potential, we impose that the $T_{cc}(3875)^+$ state appears in the vacuum as a pole in the first Riemann sheet of the BSE amplitude with
\be
\label{e:TmatrixConditions}
{\cal T}^{-1}(m_0^2) = 0, \qquad \frac{d{\cal T}^{-1}(s)}{ds}\Big |_{s=m^2_0} = \frac{1}{g^2_0} =-\frac1{P_0}\, \left. \frac{\partial\Sigma_0(s) }{\partial s}\right|_{s=m_0^2} ,
\ee
where $m_0$ is the mass of the $T_{cc}(3875)^+$. We take its value to be $0.8$ MeV below the isospin-symmetric $DD^\ast$ threshold, consistent with the analysis of Ref.~\cite{Albaladejo:2021vln}. In the last condition~\cite{Gamermann:2009uq}, we relate the coupling $g_0$ of the two-hadron pair with the derivative of the loop function at the pole position and the molecular probability content $P_0$~\cite{Weinberg:1965zz} (see also Ref.\,\cite{Albaladejo:2022sux}). Therefore, we obtain expressions for $V_A(s)$ and $V_B(s)$ in terms of $m_0$, $\Sigma_0(m_0^2)$ and the derivative $\Sigma^\prime_0(m_0^2)$. 

Similarly, we can obtain the $V$-potentials, $T$-matrices and $D^\ast D^\ast$ loop functions for the $T_{cc}(4016)^{*+}$ state, the heavy-quark partner of the $T_{cc}(3875)^+$, by substituting $D$ with $D^*$ in the expressions above. Due to HQSS, the potential remains largely the same except for some minor HQSS-breaking corrections. The $D^\ast D^\ast$ loop function then defines this potential by fixing the $T_{cc}^\ast$ mass and coupling (or molecular probability) in the vacuum. However, since the $T_{cc}^\ast$ state has not been experimentally confirmed, we take a conservative range of binding energies in the range $[0.8, 2.0]$~MeV to account for potential HQSS-breaking scenarios in our predictions. We note that, while the $DD^*$ and $D^*D^*$ channels can, in principle, mix in the $I(J^P)=0(1^+)$ sector, we neglect the coupled-channel dynamics in this work, as the coupling between these channels is expected to be suppressed in the successful hidden-gauge formalism~\cite{Dias:2021upl,Xiao:2013yca}. An assessment of the systematic uncertainties arising from this omission in the thermal modification of the $T_{cc}$ and $T_{cc}^*$ is performed in App.~\ref{ap:coupledchannels}.

Once the LECs of the potential have been determined, we compute the finite-temperature scattering amplitude by considering temperature corrections on the $D^{(\ast)}D^*$ loop function. We consider again the BS equation for the $T$-matrix at temperature $T$ as
\begin{equation}\label{eq:FTTmatrix}
    {\cal T}^{-1}(E,\vec P;T) = V^{-1}(s)-\Sigma(E, \vec P; T) ,
\end{equation}
where $\Sigma(E, \vec P; T)$ is the temperature-dependent $D^{(\ast)}D^*$ loop and $s=E^2-\vec P^2$. The interaction kernel $V(s)$ is taken to be the same as in free space. In  App.~\ref{ap:temp} we assess the impact of introducing a temperature dependence in the potential itself.

In order to compute the finite-temperature corrections to the $D^{\ast}D^{(\ast)}$ loop functions, we make use of the ITF to obtain the expression for the thermal propagator as
\begin{multline}\label{e:FTLoop}
    \Sigma(E,\vec P;T)=\int \!\! \frac{d^3q}{(2\pi)^3} \! \int_0^\infty \!\!\!\!\! d\omega \! \int_0^\infty \!\!\!\!\! d\omega^\prime \Bigg\{ 
    \big[1+f(\omega,T)+f(\omega^\prime,T)\big] \left[\frac{S_{D^{(\ast)}}(\omega,\vec q\,)S_{D^\ast}(\omega^\prime,\vec P - \vec q\,)}{E-\omega-\omega^\prime +i\varepsilon} - \frac{S_{\olsi{D}{}^{(\ast)}}(\omega,\vec q\,)S_{\olsi D{}^\ast}(\omega^\prime,\vec P - \vec q\,)}{E+\omega+\omega^\prime +i\varepsilon}\right] \\
    + \big[f(\omega,T)-f(\omega^\prime,T)\big]
    \left[\frac{S_{\olsi{D}{}^{(\ast)}}(\omega,\vec q\,)S_{D^\ast}(\omega^\prime,\vec P-\vec q\,)}{E + \omega - \omega^\prime + i\varepsilon}-\frac{S_{D^{(\ast)}}(\omega,\vec q\,)S_{\olsi D{}^\ast}(\omega^\prime,\vec P - \vec q\,)}{E - \omega + \omega^\prime + i\varepsilon}\right]\Bigg\} ,
\end{multline}
where $f(\omega,T)=1/(e^{\omega/T}-1)$ is the Bose-Einstein distribution factor that appears from the summation over Matsubara frequencies within the ITF. The spectral functions $S_{D^{(\ast)}}(\omega,\vec{q};T)$ consider the dressing of the heavy meson by the thermal medium. In this case, we employ the spectral functions for the $D$ and $D^*$ mesons obtained in Refs.~\cite{Montana:2020lfi,Montana:2020vjg} from the interaction of the heavy mesons with a thermal bath of pions. We note that the spectral functions at finite temperature of $D^{(\ast)}$ and $\olsi{D}{}^{(\ast)}$ are the same, given that the $D^{(\ast)}\pi$ and $\olsi D{}^{(\ast)}\pi$ interactions are identical in the isospin limit. As can be observed in the left panels of Figs. 6 and 7 of Ref.~\cite{Montana:2020vjg}, the thermal bath  produces a large broadening of the $D$ and $D^*$ spectral functions already at $T=80$ MeV. We compute the $D^{(\ast)}D^*$ loop function at $\vec P=0$. Note also that the imaginary part is not UV divergent, whereas the real part is cutoff-dependent.

Having determined the loop function at finite temperature and, hence, the unitarized thermal scattering amplitude ${\cal T}(E,\vec P;T)$, the corresponding spectral function at finite temperature can be defined as
\begin{equation}
\label{eq:spec_def}
{\cal S}(E,\vec{P}=0;T)=-\frac{1}{\pi}\,{\rm Im\,} {\cal T}(E,\vec{P}=0;T) \ .
\end{equation}

For hadronic ground states, the spectral function is typically defined from the imaginary part of the propagator. However, this definition becomes problematic for purely dynamically generated states, such as the $T_{cc}^{(*)}$, for which no preexisting asymptotic state exists. To address this, we have employed instead the imaginary part of the $T$--matrix for $D^{(\ast)}D^*$ scattering as a proxy for the $T_{cc}^{(*)}$ spectral function, as was done in the work of Ref.~\cite{Montana:2022inz}. The relation between the $T$--matrix and the self-energy formalisms was shown in Ref.~\cite{Albaladejo:2021cxj} for the nuclear medium and is analogously applicable to the thermal medium. In particular, it was shown that when considering a potential of the type of $V_B$, which includes an explicit exchange of a bare $T_{cc}^{(*)}$ pole, the definition of the spectral function stemming from Eq.~\eqref{eq:spec_def} is equivalent to the usual definition in terms of the renormalized propagator, up to a normalization factor. Conversely, for a $V_A$-type potential, this equivalence holds exactly only in the limit $P_0\to1$, which would correspond to the pure hadronic molecular case, and is only approximate in the region around the $T_{cc}$ vacuum mass for smaller molecular components (see Ref.~\cite{Albaladejo:2021cxj} for more details).

\section{Results}
\label{sec:results}

In this section, we present the thermal evolution of the $T_{cc}(3875)^+$ and its HQSS partner from the finite-temperature formalism described above. We first show the temperature dependence of the $D^{(\ast)}D^\ast$ loop function, and then the resulting modifications of the spectral function for different molecular components. In our approach, only the molecular component is dressed by the medium, while the compact tetraquark piece is kept unchanged. This choice is justified because a short-distance, compact configuration is expected to be less sensitive to the hadronic medium at low temperatures and may persist well above the deconfinement transition, in contrast to the extended molecular component~\cite{Esposito:2020ywk,Wu:2020zbx}.
Other sources of systematic uncertainties, namely coupled-channel dynamics and the possible thermal dependence of the interaction kernel, are examined in App.~\ref{ap:coupledchannels} and \ref{ap:temp}, respectively. Both effects lead to only mild modification of the results presented below.

\subsection{\boldmath The \texorpdfstring{$T_{cc}(3875)^+$}{Tcc(3875)+} state}
\label{ss:resultsTcc}

\begin{figure}[ht]
    \centering
    \includegraphics[width=0.7\linewidth]{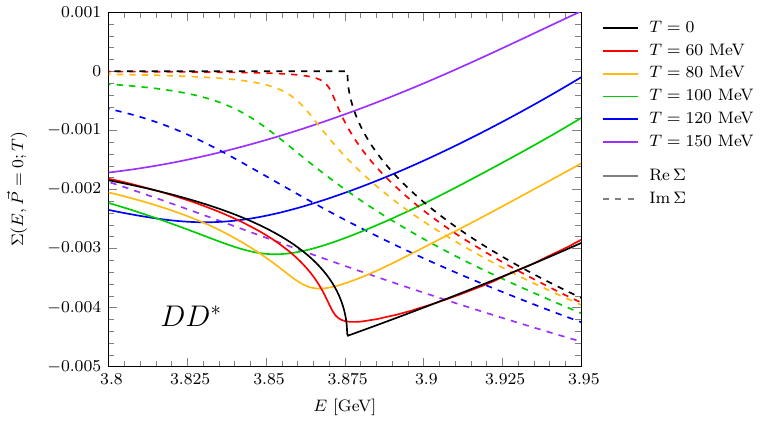}
    \caption{Real (solid) and imaginary (dashed) parts of the $DD^\ast$ loop function at several temperatures, ranging from $T=0$~MeV to $T=$150~MeV.}
    \label{fig:SigmaDDstar}
\end{figure}

We start by showing the thermal $DD^\ast$ loop function at finite temperature in Fig.~\ref{fig:SigmaDDstar}. The real part of the loop function is calculated from Eq.\,\eqref{e:FTLoop} with a cutoff regularization of $\Lambda=0.7\,\text{GeV}$, as previously mentioned, while the imaginary part follows directly from the residues of the poles in the integrand. For the computation of the real and imaginary parts of the $DD^\ast$ loop function at finite temperature, we have used the spectral functions of the $D$ and $D^*$ obtained in Refs.\,\cite{Montana:2020lfi,Montana:2020vjg}, as explained in Sec.\,\ref{sec:formalism}. In these references, it was determined that the thermal masses decrease as the temperature increases, while their widths grow substantially with temperature. Apart from the mass shift and the broadening, the spectral functions show little structure, especially when compared with the rich behavior exhibited by the finite nuclear-density ones used in Ref.~\cite{Montesinos:2023qbx} (see Figs. 6 and 7 of Ref.~\cite{Montana:2020vjg}).

To interpret the impact of the thermal dressing of the $DD^\ast$ loop on the scattering amplitude, and following Refs.~\cite{Montesinos:2023qbx,Montesinos:2024xpp}, it is convenient to rewrite the BSE of Eq.~\eqref{eq:FTTmatrix} as (ommitting the $\vec P=\vec 0$ label) 
\begin{equation}\label{eq:BSEVeff}%
\mathcal{T}^{-1}(E;T) = V_{\rm eff}^{-1}(E;T) -\Sigma_0(s)\,,
\end{equation}
where $\Sigma_0$ is the $DD^\ast$ loop in the free space and $V_\mathrm{eff}$ is an effective interaction incorporating the thermal effects,
\begin{subequations}\label{e:Veff}
    \begin{align}
    V_{\rm eff}^{-1}(E;T) & = V^{-1}(s)+\delta\Sigma(E;T)\,,\\
    \delta\Sigma(E;T) & =\Sigma_0(s)-\Sigma(E;T)\,.
    \end{align}
\end{subequations}

As a consequence of the thermal modifications to the $D^{(\ast)}$ mesons, the dressing of the loop functions with the $D^{(\ast)}$ spectral functions softens and shifts towards lower energies the onset of the unitary cut of the imaginary part with increasing temperature (dashed lines). The real part (solid lines) changes with temperature accordingly, with the cusp at the $DD^*$ threshold lowering at larger temperatures. The real part also becomes less negative with increasing temperature for energies around the threshold. Since $\mathrm{Re}~ \delta\Sigma(E;T) <0$, we have from Eqs.~\eqref{e:Veff} 
that the effective interaction becomes more repulsive ($\mathrm{Re}~V_\mathrm{eff}>V$), although this interpretation must be taken with caution due to the sizeable imaginary part of the thermal loop developing below threshold.  We note that the behavior of the imaginary part of the $DD^*$ loop with energy and temperature is the same as the one reported for $D \olsi{D}{}^{\ast}$ in Fig.~2 of Ref.~\cite{Montana:2022inz}, as expected. The real part behaves similarly with energy and temperature, although the exact values depend on the chosen regularization scale, which was different in the work of Ref.~\cite{Montana:2022inz}.

\begin{figure}[ht]
    \centering
    \includegraphics[width=0.45\linewidth]{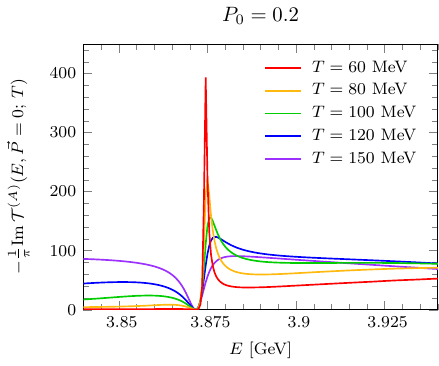}
    \includegraphics[width=0.45\linewidth]{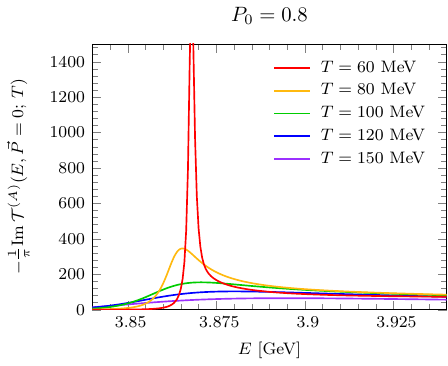}
    \includegraphics[width=0.45\linewidth]{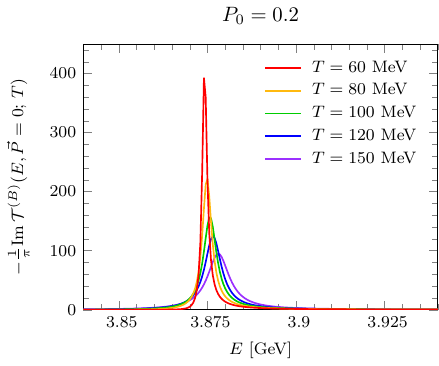}
    \includegraphics[width=0.45\linewidth]{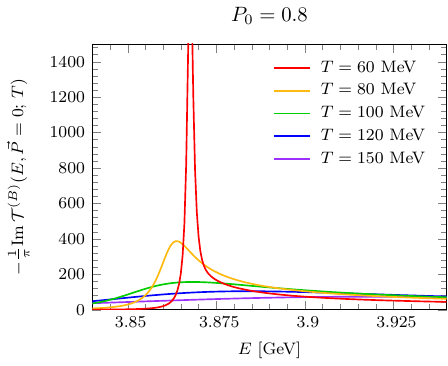}
    \caption{$T_{cc}(3875)$ spectral functions as a function of energy for different temperatures, computed using $V_A$ [top row, \textit{cf.} Eq.\,\eqref{eq:VA}] and $V_B$ [bottom row, \textit{cf.} Eq.\,\eqref{eq:VB}], for two values of the molecular probability (left and right panels, respectively).}
    \label{fig:TccSpectralFunction}
\end{figure}

Once we have obtained the temperature-dependent $DD^*$ loop functions $\Sigma(E,\vec{P};T)$, the $DD^*$ $T$-matrices and the corresponding $T_{cc}(3875)^+$ spectral functions ($\mathcal{S}_{T_{cc}}$) at finite temperature can be calculated from Eqs.~\eqref{eq:FTTmatrix} and \eqref{eq:spec_def}, respectively. In Fig.~\ref{fig:TccSpectralFunction} we show the spectral functions considering the two families of potentials, $V_A$ (upper row) and $V_B$ (lower row) for molecular probabilities $P_0 = 0.2$ (left column) and $0.8$ (right column). For each plot, we show the $\mathcal{S}_{T_{cc}}$ for different temperatures, ranging from $T=60$~MeV to $T=150$~MeV, close to the temperature of the QCD phase transition to quark matter.

If we compare the $\mathcal{S}_{T_{cc}}$ computed using the $V_A$ potential and the ones obtained from the $V_B$ potential, we find that for high values of the molecular $DD^*$ component (right column) the results for both potentials are almost identical for all temperatures. As discussed in Refs.~\cite{Albaladejo:2021cxj,Montesinos:2023qbx}, the zero of $V_A$ and the bare pole of $V_B$ are far from the energies considered to induce any significant changes. For the small value of $P_0=0.2$ (left column), both potentials show significant deviations from each other, leading to different $T$-matrices at finite temperature, although both potentials give the same $T_{cc}(3875)^+$ mass and $DD^{\ast}$ coupling at zero temperature. A similar behavior was observed when the  $T_{cc}(3875)^+$ exotic state was embedded in a dense medium~\cite{Montesinos:2023qbx}.

As for the temperature dependence for small and large $P_0$, we find that the spectral functions $\mathcal{S}_{T_{cc}}$ change more rapidly with temperature for high molecular probabilities. For large values of $P_0$ (right column), the width increases significantly with temperature, leading to the melting of the $T_{cc}(3875)^+$ for temperatures larger than 80 MeV. As for a small molecular component of $P_0=0.2$ (left column), the changes in the spectral function due to temperature become less important, differing according to the potential used, as already indicated before. More precisely, the spectral functions obtained with $V_A$ show the zero that this type of potential has, and vary with temperature above and below this zero, as already seen in the case of finite density in Ref.\,\cite{Montesinos:2023qbx}. For the case of the $V_B$ interaction, the quasi-particle peak, induced by the bare pole present in the potential, moves to higher energies while slowly melting down with temperature, in contrast to the case of large $P_0$. All these results indicate that any experimental determination of the scattering amplitudes at finite temperature can provide valuable insights into the molecular structure of the $T_{cc}(3875)^+$ exotic state.

\subsection{\boldmath The HQSS partner of the \texorpdfstring{$T_{cc}(3875)^+$}{Tcc(3875)+}: the \texorpdfstring{$T_{cc}(4016)^{*+}$}{Tcc(4016){*+}} state}
\label{ss:resultsheavy-quark}

\begin{figure}[ht]
    \centering
    \includegraphics[width=0.7\linewidth]{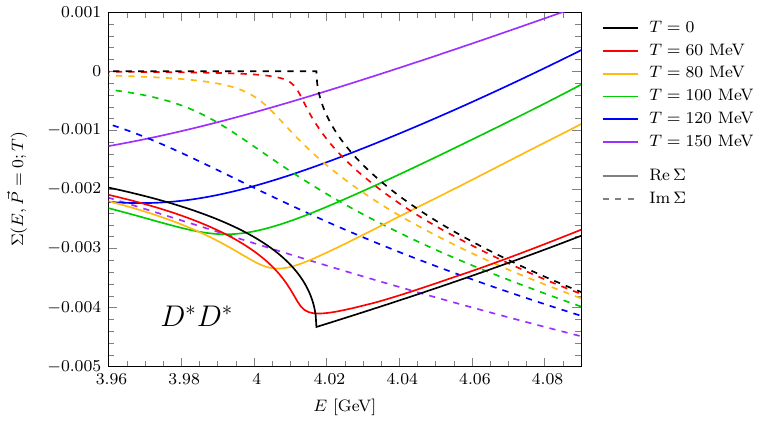}
    \caption{Real (solid) and imaginary (dashed) parts of the $D^\ast D^\ast$ loop function at several temperatures, ranging from $T=0$~MeV to $T=$150~MeV.}
    \label{fig:SigmaDstarDstar}
\end{figure}

In Fig.~\ref{fig:SigmaDstarDstar} we present the results for the $D^\ast D^\ast$ loop function. Similarly to the case of the $DD^\ast$ loop, we also use a cutoff regulator $\Lambda = 0.7$ GeV for the evaluation of its real part. The obtained results are qualitatively similar to those of the $DD^\ast$ loop, but shifting the opening of the unitarity cut by the appropriate mass difference $m_{D^\ast}-m_D\approx m_\pi$. As in the $DD^\ast$ case, we can observe a general softening and shift towards lower energies of the onset of the unitarity cut, in both the real (solid lines) and imaginary (dashed lines) parts. This is again due to the decrease in the thermal mass of the $D^\ast$ mesons found in Refs.~\cite{Montana:2020lfi,Montana:2020vjg}. One can observe in the $D^\ast D^\ast$ case a slightly less attractive real part, as compared with the real part of the $DD^\ast$ loop. Conversely, the imaginary part is also slightly larger. However, these differences are not very notable.

\begin{figure}[ht]
    \centering
    \includegraphics[width=0.45\linewidth]{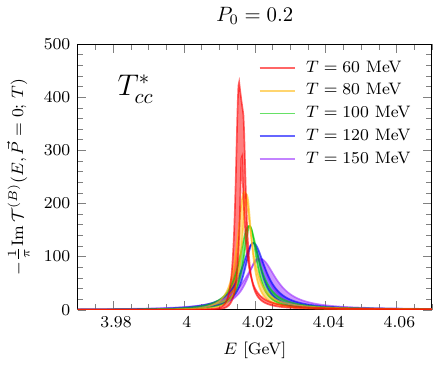}
    \hfill
    \includegraphics[width=0.45\linewidth]{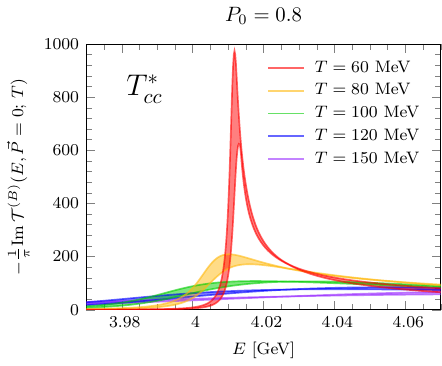}
    \caption{ $T_{cc}(4016)^{\ast+}$ spectral functions as a function of energy for different temperatures, computed using and $V_B$, for two values of the molecular probability (left and right panels, respectively). The shaded bands correspond to considering the $T_{cc}^\ast$ binding energy w.r.t. the $D^\ast D^\ast$ threshold in the interval $[0.8, 2.0]$ MeV.}
    \label{fig:TccstarSpectralFunction}
\end{figure}

Given that the $I(J^P)=0(1^+)$ $DD^\ast$ and $D^\ast D^\ast$ interaction kernels are the same due to HQSS, the $T_{cc}^\ast$ state is predicted to have a binding energy similar to that of the $T_{cc}$, now taken with respect to the $D^\ast D^\ast$ threshold. Actually, the $T_{cc}^\ast$ is expected to be slightly more bound than the $T_{cc}$, owing to the modification of the kinetic term of the $D^\ast D^\ast$ channel. In our approach, we take a range of values for the binding energy, which is defined as 
\begin{equation}
    E_b=2 m_{D^\ast}-m_0^{T_{cc}^\ast},
\end{equation} 
so as to analyze the effect of this parameter on the lineshape of the $T_{cc}^\ast$ spectral function $\mathcal{S}_{T_{cc}^\ast}$. The results for this function can be found in the panels of Fig.~\ref{fig:TccstarSpectralFunction}. We plot $\mathcal{S}_{T_{cc}^\ast}$ derived from the $V_B$ interaction, for several values of the temperature, and considering the same two molecular probabilities as in the case of the $T_{cc}$. A comparison with the results obtained when using the $V_A$ potential will be redundant with what has already been discussed for the $T_{cc}$. The shaded bands include the different solutions obtained when varying the binding energy $E_b$ in the interval $[0.8,2.0]$ MeV. In Ref.~\cite{Albaladejo:2021vln}, the binding of the $T_{cc}^\ast$ with respect to the $D^{\ast +} D^{\ast 0}$ threshold was predicted to be around $1.2 \sim 1.6$ MeV, depending on the regularization scale chosen. These values are included in the range of binding energies that we consider.

When comparing the results obtained here for the $T_{cc}^\ast$ with those shown in Fig.\,\ref{fig:TccSpectralFunction} for the $T_{cc}$, we observe a similar qualitative behavior. This was also the case in Ref.\,\cite{Montana:2022inz}, when comparing the temperature effects on the $X(3872)$ and its $2^{++}$ heavy quark spin partner $X(4014)$ (see Fig.~3 in that reference). In the study of Ref.\,\cite{Montesinos:2023qbx} of the $T_{cc}$ and $T_{cc}^\ast$ in the presence of a dense nuclear medium, a similar qualitative behavior with density was also found for the two HQSS partners. Regarding the different binding-energy scenarios considered, which we interpret as an estimate of the systematic uncertainties of the HQSS approach, only moderate effects are observed compared to the large thermal effects. For $P_0=0.8$, the $T_{cc}^\ast$ is consistently melted at temperatures around $100$~MeV independently of the binding energy. A comparison of the two panels of Fig.~\ref{fig:TccstarSpectralFunction} shows again the high sensitivity of the spectral function to the molecular probability: its thermal evolution is significantly milder in the low molecular probability scenario than in the high molecular probability case.

\section{Conclusions}
\label{sec:conclusions}
We have discussed the spectral properties of the tetraquark-like $T_{cc}(3875)^+$ and $T_{cc}(4016)^{*+}$ immersed in a hot bath of pions. We have considered these exotic states as purely isoscalar $DD^\ast$ and $D^\ast D^\ast$ $S$-wave bound states, respectively. For the BSE interaction kernel at zero temperature, we have considered two families of energy-dependent interactions that allow for a more exhaustive analysis of the molecular probability content of these states. The effects of finite temperature are incorporated up to $T=150$ MeV through the $D^{(\ast)}$ thermal spectral functions calculated in Ref.~\cite{Montana:2020vjg}, with the changes in the $D^{(\ast)}$ propagators implemented using the ITF.

We have found important modifications of the $D^{\ast}D^{(\ast)}$ scattering amplitudes already for $T=80$ MeV. They are produced by the dressing of the two open-charm-meson loop functions with the $D$ and $D^*$ spectral functions, which softens and shifts towards lower energies the onset of the unitary cut of the imaginary part with increasing temperature.  The real part changes with temperature accordingly,
with the cusp at the $DD^*$ threshold lowering at larger temperatures. Next, we have discussed the dependence of the hot-bath lineshapes of these tetraquark-like states on their Weinberg molecular content, and found that the $\mathcal{S}_{T^{(\ast)}_{cc}}$ spectral functions change more rapidly with temperature for high molecular probabilities. For large values of $P_0$, the widths significantly increase with temperature, leading to the melting of these exotic states for temperatures larger than $100$~MeV. For small molecular components, the changes in the spectral functions of these states due to temperature become less important, and depend significantly on the employed BSE potential kernel, although these results should be interpreted with caution, as only the molecular component is affected by the medium in our approach. The compact tetraquark component is expected to be less sensitive to a low-temperature hadronic medium and may survive well above the deconfinement transition, so the qualitative signature of molecular-content dependence in the thermal evolution of the $T_{cc}$ and $T_{cc}^\ast$ is expected to remain robust.


These results show that any experimental determination of the $D^{(\ast)}D^*$ scattering amplitudes at finite temperature, in experiments like RHIC or LHC, will provide valuable insights into the molecular structure of the $T_{cc}(3875)^+$ and $T_{cc}(4016)^{*+}$ states. Furthermore, by combining the thermal line-shape patterns of these states with future measurements at FAIR (CBM, PANDA) of their spectral properties in dense nuclear environments~\cite{Montesinos:2023qbx}, as well as those of their antiparticles, it should be possible to very efficiently constrain the nature of these enigmatic tetraquark-like exotics. In addition, the in-medium scattering amplitudes and spectral functions obtained in our framework may be used in simulations of heavy-ion collisions. A natural extension of our study of the $T_{cc}$ spectral function is the computation of $T_{cc}$ Euclidean correlators at finite temperature, which would allow comparison with future lattice QCD results, and offer further insight into its in-medium properties and molecular content.

\section*{Acknowledgments}
We acknowledge support from the programs Unidades de Excelencia Severo Ochoa CEX2023-001292-S and María de Maeztu CEX2020-001058-M, from the projects PID 2020-112777GB-I00, PID2022-139427NB-I00 and PID2023-147458NB-C21  financed by the Spanish MCIN/AEI/10.13039/501100011033, 
as well as from the Grant CIPROM 2023/59 of Generalitat Valenciana, the Generalitat de Catalunya under contract 2021 SGR 171, and by the CRC-TR 211 ’Strong-interaction matter under extreme conditions’-project Nr. 315477589 - TRR 211. %
M.\,A. acknowledges financial support through GenT program by Generalitat Valencia (GVA) Grant No.\,CIDEGENT/2020/002, Ramón y Cajal program by MICINN Grant No.\,RYC2022-038524-I, and ``Atracción de Talento'' program by CSIC, Grant No. PIE 20245AT019. %
G.M. was supported by the U.S. Department of Energy contract \mbox{DE-AC05-06OR23177}, under which Jefferson Science Associates, LLC operates Jefferson Lab. V.M. acknowledges support by GVA under Grant No. ACIF/2021/290.

\appendix

\section{Impact of coupled-channel dynamics}\label{ap:coupledchannels}

In this work, the $DD^\ast$ and $D^\ast D^\ast$ $I(J^P)=0(1^+)$ channels were treated as independent, and the $T_{cc}$ and $T_{cc}^\ast$ were analyzed separately. However, since both share the same quantum numbers ($I(J^{P})=0(1^+)$), these channels are necessarily coupled. In the following, we estimate the effect of the coupled-channel dynamics on the results presented in Sec.~\ref{sec:results}.

We consider the following potential consistent with HQSS,
\begin{equation}
    \hat V =
    \begin{pmatrix}\label{e:CoupledChannelsPotential}
        v &\xi v \\
        \xi v & v
    \end{pmatrix} \ ,
\end{equation}
in the basis $\{\ket{DD^\ast},\ket{D^\ast D^\ast}\}$. We shall take the diagonal matrix element $v$ as
\begin{equation}\label{e:CoupledChannelsDiagonalTerm}
    v = \frac{1}{\Sigma_0(m_0)}=\text{const.} \ ,
\end{equation}
where $\Sigma_0$ denotes the $DD^\ast$ loop in free space, computed with a cutoff regulator $\Lambda=0.7$ GeV, and $m_0$ is the mass of the $T_{cc}$ in vacuum. 

The off-diagonal elements contain a parameter $\xi$ accounting for the strength of the coupling between the two channels. For $\xi=0$, the system reduces to the decoupled case discussed previously, for purely molecular states ($P_0=1$). For $\xi>0$, both the $T_{cc}$ and $T_{cc}^\ast$ acquire admixtures of $\ket{DD^\ast}$ and $\ket{D^\ast D^\ast}$ components. The value of $\xi$ is expected to be small, since in the successful hidden gauge formalism~\cite{Xiao:2013yca}, the diagonal $DD^\ast \to DD^\ast$ and $D^\ast D^\ast \to D^\ast D^\ast$ transitions are dominated by $\rho$ exchange in $S$--wave, which is forbidden for the inelastic transition $D^\ast D^\ast\to DD^\ast$. The latter proceeds through pion and rho exchanges in $P$--wave and is therefore suppressed~\cite{Dias:2021upl}. We thus consider $\xi\leq 0.3$.

The coupled-channels BSE takes a matrix form, 
\begin{equation}
    \hat{\mathcal{T}}(E;T) = \left[\mathbb{I}_{2\times2}-\hat V \hat \Sigma(E;T)\right]^{-1} \hat V,
\end{equation}
where the loop-function matrix is diagonal,
\begin{equation}
    \hat \Sigma(E;T)=
    \begin{pmatrix}
        \Sigma_{DD^\ast}(E;T) & 0 \\
        0 & \Sigma_{D^\ast D^\ast}(E;T)
    \end{pmatrix}
\end{equation}
and contains the thermal loop functions for the $DD^\ast$ and $D^\ast D^\ast$ systems at zero center-of-mass momentum, as given in Eq.~\eqref{e:FTLoop}.

\begin{figure}[ht]
    \centering
    \includegraphics[width=0.45\linewidth]{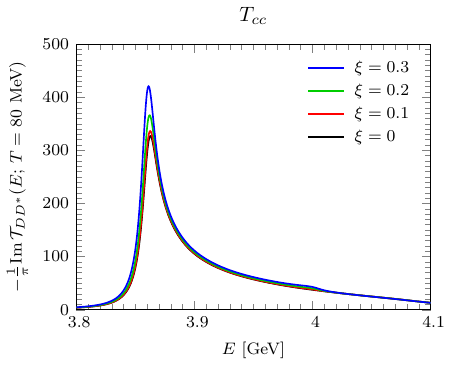}
    \hfill
    \includegraphics[width=0.45\linewidth]{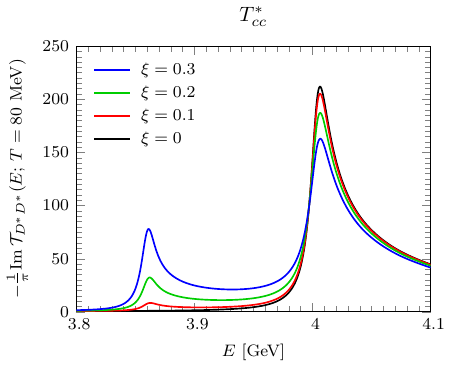}
    \caption{Illustration of the coupled-channel effects on the temperature-dependent $T_{cc}$ (left panel) and $T_{cc}^\ast$ (right panel) spectral functions, for different strengths of the coupling-channel potential parameter $\xi$, defined in Eq.~\eqref{e:CoupledChannelsPotential}.}
    \label{fig:CoupledChannels}
\end{figure}

Figure~\ref{fig:CoupledChannels} shows the imaginary parts of $\mathcal{T}_{DD^\ast}\equiv \bra{DD^\ast}\hat{\mathcal{T}} \ket{DD^\ast}$ and $\mathcal{T_{D^\ast D^\ast}}\equiv \bra{D^\ast D^\ast}\hat{\mathcal{T}} \ket{D^\ast D^\ast}$ at $T=80$ MeV for several values of $\xi$.
We select a moderate temperature of $80$ MeV for this comparison to avoid sizeable systematic uncertainties that may arise at higher temperatures due to the omission of the kaonic thermal bath and the onset of deconfinement, among others.
The $T_{cc}$ lineshape (left panel) exhibits only a slight narrowing as the coupling to the $D^\ast D^\ast$ channel increases, while the $T_{cc}^\ast$ lineshape (right panel) broadens with growing $\xi$ and mixes with the $T_{cc}$ peak. In both cases, the modifications are minor and do not significantly alter the peak positions or widths, remaining much smaller than the thermal effects and the molecular probability dependence discussed in the main text.

\section{Impact of a temperature-dependent potential}\label{ap:temp}

A possible source of systematic uncertainty arises from a temperature dependence of the interaction kernel, which was neglected in the previous analysis. Here we estimate its effect on the thermal amplitudes within the coupled-channel framework of App.~\ref{ap:coupledchannels}.

In the local hidden gauge approach~\cite{Feijoo:2021ppq}, the temperature dependence of the interaction kernel is governed by
\begin{equation}
    v\sim \frac{1}{f_\pi^2(T)}\frac{m_V^2}{m_\rho^2(T)}m_{D^{(\ast)}}^2(T) \ ,
\end{equation}
where $f_\pi$ is the pion decay constant, $m_V=0.8$~GeV is a mass parameter, and $m_\rho$  and $m_{D^{(\ast)}}$ are the $\rho$ and charmed meson masses, respectively. Assuming that the dominant thermal effect arises from the pion decay constant, we approximate
\begin{equation}
    v(T)=v\frac{f_\pi^2(0)}{f_\pi^2(T)} \ ,
\end{equation}
with $v$ defined in Eq.~\eqref{e:CoupledChannelsDiagonalTerm}. 

The temperature dependence of $f_\pi$ has been determined in several works~\cite{Gasser:1986vb,Kodama:1995kj,Bochkarev:1995gi,Jeon:1996gn,Pisarski:1996mt,Harada:1996pg}. Here, we adopt the low-temperature result from leading-order chiral perturbation theory of Ref.~\cite{Gasser:1986vb}, which was shown to coincide, for $N_f=2$ flavors, with the corresponding low-energy behavior obtained in both the $O(N)$ linear and nonlinear sigma models for $N=4$~\cite{Bochkarev:1995gi}:
\begin{align}
    \frac{f_\pi(T)}{f_\pi(0)}&\approx 1-\frac{T^2}{12f_\pi^2(0)} \ . \label{e:fpiChPTLO} 
\end{align}

\begin{figure}[ht]
    \centering
    \includegraphics[width=0.45\linewidth]{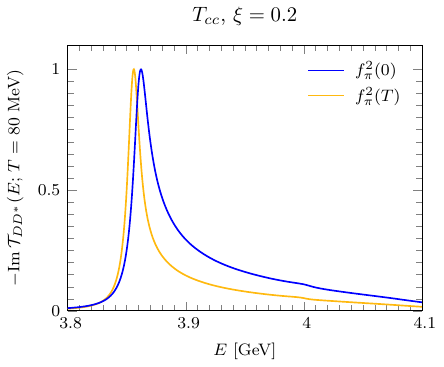}
    \hfill
    \includegraphics[width=0.45\linewidth]{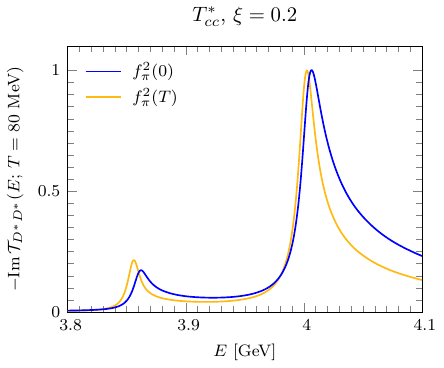}
    \caption{ Comparison of the coupled-channel results for the $T_{cc}$ and $T_{cc}^\ast$ spectral functions when considering or not the temperature effects in the interaction kernel. The temperature has been taken as $T=80$ MeV, and the $\xi$ parameter of Eq.~\eqref{e:CoupledChannelsPotential} is assumed to be $0.2$. The results have been normalized to unity at their respective maxima.}
    \label{fig:CoupledChannelsTemp}
\end{figure}

Figure~\ref{fig:CoupledChannelsTemp} shows the $T_{cc}$ (left) and $T_{cc}^\ast$ (right) spectral functions for $T=80$ MeV and $\xi=0.2$, while comparing the results obtained when including the temperature dependence of the potential (orange lines) and when not considering it (blue lines, equivalent to green lines in Fig.~\ref{fig:CoupledChannels}), normalized to unity at their respective maxima. The inclusion of the temperature dependence produces a shift of the quasiparticle masses towards lower energies and a reduction of their widths. This behavior is related to the reduction of the thermal pion decay constant with increasing temperature, which produces a more attractive potential. These results should, however, be interpreted with caution, since the potential is closely related to the renormalization scale of the loop function, i.e., the value of the cutoff (which we have fixed to $\Lambda=0.7$ GeV), which may also depend on the temperature. This cutoff is already difficult to fix in the free space for coupled channels~\cite{Nieves:2012tt}. In addition, it should be noted that the $D$ and $D^*$ spectral functions employed here were obtained assuming a temperature-independent interaction with the thermal bath, so the present implementation of a thermal dependence in the $D^{(*)}D^*$ interaction should be regarded only as an estimate of the expected magnitude of such effects rather than a fully consistent treatment.

\bibliographystyle{JHEP}
\bibliography{references.bib}

\end{document}